\documentclass[reprint,prd,twocolumn,superscriptaddress,showpacs,showkeys,floatfix,preprintnumbers]{revtex4-2}
\usepackage{mathrsfs}
\usepackage{amsfonts}
\usepackage{amsmath}
\usepackage{slashed}
\usepackage{pifont}
\usepackage{array}
\usepackage{verbatim}
\usepackage{epsfig}
\usepackage{graphicx}
\usepackage{booktabs}
\usepackage{subfigure}
\usepackage{float}
\usepackage{multirow}
\usepackage{color}
\usepackage[dvipsnames]{xcolor}
\definecolor{ceruleanblue}{rgb}{0.16, 0.32, 0.75}
\usepackage[colorlinks=true,linkcolor=ceruleanblue,citecolor=ceruleanblue,urlcolor=blue]{hyperref}

\newcommand{\beq}{\begin{eqnarray}}
\newcommand{\eeq}{\end{eqnarray}}
\newcommand{\be}{\begin{equation}}
\newcommand{\ee}{\end{equation}}
\newcommand{\ba}{\begin{eqnarray}}
\newcommand{\ea}{\end{eqnarray}}

\newcommand{\bit}{\begin{itemize}}
\newcommand{\eit}{\end{itemize}}
\newcommand{\rw}{\rightarrow}

\newcommand{\zzuphy}{School of Physics, Zhengzhou University, Zhengzhou, Henan 450001, China}
\newcommand{\innovation}{Collaborative Innovation Center of Quantum Matter, Beijing 100871, China}
\newcommand{\chep}{Center for High Energy Physics, Peking University, Beijing 100871, China}
\newcommand{\pkuphy}{School of Physics, Peking University, Beijing 100871,
China}

\begin{document}

\title{Lattice study of $J/\psi \rw \gamma\eta_c$ using a method without momentum extrapolation}

\author{Yu Meng}
\email[Email: ]{yu\_meng@zzu.edu.cn}
\affiliation{\zzuphy}
\author{Chuan Liu}
\email[Email: ]{liuchuan@pku.edu.cn}
\affiliation{\pkuphy}\affiliation{\chep}\affiliation{\innovation}
\author{Teng Wang}\affiliation{\pkuphy}
\author{Haobo Yan}\affiliation{\pkuphy}

\date{\today}

\begin{abstract}
We present a model-independent method to calculate the radiative transition without the momentum extrapolation for the off shell transition factors. The on shell transition factor is directly obtained from the lattice hadronic function. We apply the method to calculate the charmonium radiative transition $J/\psi \rw \gamma\eta_c$. After a continuous extrapolation under three lattice spacings, we obtain the on shell transition 
factor as $V(0)=1.90(4)$, where the error is the statistical error that already takes into account the $a^2$ error in the continuous extrapolation. Finally, we determine the branching fraction of $J/\psi\rw \gamma \eta_c$ as $\operatorname{Br}(J/\psi\rightarrow \gamma\eta_c)=2.49(11)_{\textrm{lat}}(5)_{\textrm{exp}}\%$, where the second error comes from the uncertainty of $J/\psi$ total decay width $92.6(1.7)$ keV.
\end{abstract}

\maketitle

\section{Introduction}
Charmonium, a bound state particle composed of a charm quark and its antiparticle, has been a subject of extensive theoretical and experimental interest since its discovery five decades ago~\cite{Jpsi1974a,Jpsi1974b}. On one hand, the world's largest charm factory---BESIII collaboration, which has collected the largest number of $J/\psi$ particles~\cite{BESIII:2021cxx}, will further improve the experimental precision of various related physical processes. On the other hand, due to charmonium's intermediate energy scale where both perturbative and nonperturbative methods are applicable, this particle offers an excellent avenue for testing various theories and methods, providing an ideal ground for a deeper understanding of the strong interaction.
The radiative transition process of charmonium, where the initial vector particle $J/\psi$ emits a real photon to transfer to the lowest pseudoscalar state, is the simplest physical process in the charmonium family. However, to date, the direct experimental measurements are very limited~\cite{Gaiser:1985ix,CLEO:2008pln,Anashin:2014wva}, and the measured branching ratios also have large uncertainties~\cite{CLEO:2008qfy,BESIII:2012lxx}.  The latest Particle Data Group(PDG) has updated the branching ratio for this process to be 1.41(14)\%~\cite{PDG24}, leading to a significant improvement in precision compared to the previous well-known 1.7(4)\%~\cite{PDG22}.

On the theoretical side, the radiative transition of charmonium involves both electromagnetic and strong interactions. Since the charmonium is in  an intermediate energy scale, various perturbative and nonperturbative methods are proposed and applied to this fundamental process. Among them, the genuine nonperturbative method such as lattice QCD has played a key role. In traditional lattice calculations~\cite{Dudek:2006ej,Chen:2011kpa,Gui:2019dtm,Delaney:2023fsc,Li:2023zig}, the on shell transition factor is determined by the extrapolation of the off shell transition factors with
nonzero photon virtualities, or by so-called twisted boundary conditions~\cite{Bedaque:2004kc,deDivitiis:2004kq} where an appropriate twisted angle is tuned
to put the transition factor on shell directly. The former inevitably leads to a model-dependent error caused by the momentum extrapolation of the 
off shell transition factors, while the latter needs to produce the specific propagators which
are usually difficult to use for other lattice calculations.

In a recent work~\cite{Feng:2019geu}, a model-independent method is
proposed to compute the pion charge radius on the lattice. Such a method was originally put forward to avoid model-dependent momentum extrapolation. The key point is to construct an appropriate scalar function and get the physical quantity by projecting a related momentum. A similar idea has been widely applied to various processes~\cite{Feng:2020zdc,Ma:2021azh,Tuo:2021ewr,Meng:2021ecs,Fu:2022fgh,Meng:2024gpd,Meng:2024nyo,Tuo:2024bhm}. In this paper, we would apply the idea to the radiative transition. However, the situation here is generally different from the calculation of charge radius, where the initial and final states are the same particles and the on shell transition factor is straightforwardly calculated by projecting zero momentum on the final state. In the case of radiative transition, the on shell transition factor does not involve the zero momentum projection of the final state anymore$-$but a particular momentum that is not directly accessible on the lattice. Therefore, we develop the method and, as an example, apply it to the calculation of charmonium transition, i.e. $J/\psi \rw \gamma\eta_c$, which has been studied comprehensively by the traditional method~\cite{Dudek:2006ej,Chen:2011kpa,Gui:2019dtm,Delaney:2023fsc,Li:2023zig} and twisted boundary conditions~\cite{Becirevic:2012dc,Donald:2012ga,Colquhoun:2023zbc}.

The rest of this paper is organized as follows. In Sec.~\ref{sec:method}, we
introduce the new methodology to calculate the on shell transition factor using only the lattice data as input. In Sec.~\ref{sec:setup}, the configuration information is given. In Sec.~\ref{sec:result}, we provide details of the simulations and show the main results. This section is further divided into three parts: in Sec.~\ref{sec:disper_relation} the dispersion relation of $\eta_c$ mass is presented; in Sec.~\ref{sec:FVC} the finite-volume effects are discussed; in Sec.~\ref{sec:form_factor} the numerical results of transition form factors are presented and the continuum limit under three lattice spacings is performed. Finally, we present the discussion on the advantages of the method in Sec.~\ref{sec:discussion} and conclude in Sec.~\ref{sec:conclude}.

\section{Methodology}\label{sec:method}
We start with a Euclidean hadronic function in the infinite volume
\be
H_{\mu\nu}(\vec{x},t)= \langle 0|\phi_{\eta_c}(\vec{x},t)J_{\nu}(0)|J/\psi_{\mu}(p')\rangle, t>0
\ee
where $|J/\psi_{\mu}(p')$ is a $J/\psi$ state with four-momentum $p'=(im_{J/\psi},\vec{0})$ and $\phi_{\eta_c}$ is the interpolating operator of $\eta_c$. $J_{\nu}$ is an electromagnetic vector current with the form of $J_{\nu}=\sum_qe_q\,\bar{q}\gamma_\mu q$ ($e_q=2/3,-1/3,-1/3,2/3$ for $q=u,d,s,c$).

At large time $t$, the hadroinc function is saturated by the single $\eta_c$ state
\beq\label{eq:H_munu}
H_{\mu\nu}(\vec{x},t)&\doteq& \frac{2e_c}{m_{\eta_c}+m_{J/\psi}}\int\frac{d^3\vec{p}}{(2\pi)^3}
\frac{Z}{E}\epsilon_{\mu\nu\alpha \beta} p_{\alpha}p'_{\beta} \nonumber \\
&\times& V(q^2)e^{-E t+i\vec{p}\cdot \vec{x}},
\eeq
where the overlap function $Z$ and transition factor $V(q^2)$
are defined by
\beq
\langle 0|\phi_{\eta_c}(0)|\eta_c(\vec{p})\rangle &=&Z \\
\langle \eta_c(\vec{p})|J_{\nu},(0)|J/\psi_{\mu}(p')\rangle&=&\frac{4V(q^2)}{m_{\eta_c}+m_{J/\psi}}e_c\epsilon_{\mu\nu\alpha \beta}p_{\alpha}p'_{\beta},
\eeq
with $q^2=(m_{J/\psi}-E)^2- |\vec{p}|^2$.
The discretization effect on the lattice not only breaks continuous dispersion relation,
but the Lorentz invariance, thus resulting in a momentum dependence of $Z$. In this paper, we will use the following parametrizations:
\beq\label{eq:E_Z}
E^2&=& m_{\eta_c}^2+\xi\cdot |\vec{p}|^2,  \nonumber \\
Z^2&=&Z_0^2+\eta\cdot |\vec{p}|^2,
\eeq
where the modified coefficients $\xi$ and $\eta$ are introduced.

To compute the $V(0)$, we construct a scalar function $\mathcal{I}_0(t,|\vec{p}|)$ by multiplying $\epsilon_{\mu\nu\alpha' \beta'}p_{\alpha'}p'_{\beta'}/(m_{J/\psi}|\vec{p}|^2)$ to the spatial Fourier transform of $H_{\mu\nu}(\vec{x},t)$, it yields
\beq\label{eq:main}
\mathcal{I}_0(t,|\vec{p}|)
&=&-\frac{4e_cZm_{J/\psi}}{m_{\eta_c}+m_{J/\psi}}V(q^2)\frac{e^{-Et}}{E} \nonumber \\
&=&\frac{1}{ |\vec{p}|^2}\int d^3\vec{x}e^{-i\vec{p}\cdot\vec{x}}\epsilon_{\mu\nu\alpha0}\frac{\partial H_{\mu\nu}(x) }{\partial x_{\alpha}}.
\eeq
After averaging over the spatial direction for $\vec{p}$,
\be\label{eq:I_0}
\mathcal{I}_0(t,|\vec{p}|)=\int d^3\vec{x}\frac{j_1(|\vec{p}||\vec{x}|)}{|\vec{p}||\vec{x}|}\epsilon_{\mu\nu \alpha 0}x_{\alpha}H_{\mu\nu}(\vec{x},t),
\ee
where $j_n(x)$ are the spherical Bessel functions. In the Taylor expansion at $q^2=0$, the transition factor has the form
\beq\label{eq:V_taylor}
V(q^2)&=&\sum\limits_{n=0}^{\infty}c_n\left(\frac{q^2}{m_{J/\psi}^2}\right)^n \nonumber \\
&\doteq & c_0+c_1\cdot \frac{q^2}{m_{J/\psi}^2}+\mathcal{O}(q^4/m_{J/\psi}^4),
\eeq
where the symbol $\doteq$ denotes the omission of high-order terms $c_{n\geq 2}$.
The $c_0$ is the on shell transition factor $c_0\equiv V(0)$, and $c_1$ is related to the slope
of the transition factor $V(q^2)$ at $q^2=0$. We can obtain $c_0$ immediately by
taking a derivative of the above equation at $|\vec{p}|^2=0$, which corresponds to
$q^2=(\delta m)^2 \equiv (m_{J/\psi}-m_{\eta_c})^2$. These high-order terms $c_{n\geq 2}$
are expected to be negligible for the calculation of $c_0$ since $(\delta m)^2/m_{J/\psi}^2 \sim 0.14\%$; the effect of $c_{n\geq 2}$ on $c_0$ is high-supressed by the factor $(0.14\%)^n$.

The derivative of $\mathcal{I}_0(t,|\vec{p}|)$ at $|\vec{p}|^2=0$, on one hand, leads to
\beq\label{eq:I_1}
\mathcal{I}_1(t,0)&\equiv& -\frac{\partial \mathcal{I}_0(t,|\vec{p}|)}{\partial |\vec{p}|^2}\Big{|}_{|\vec{p}|^2=0} \nonumber \\
&=&\frac{1}{30}\int d^3\vec{x}|\vec{x}|^2\epsilon_{\mu\nu\alpha 0}x_{\alpha}H_{\mu\nu}(\vec{x},t),
\eeq
on the other hand, it has
\beq
&&\mathcal{I}_1(t,0) \nonumber \\
&=&\frac{-4e_c\xi Z_0 m_{J/\psi}}{m_{\eta_c}+m_{J/\psi}}\frac{e^{-m_{\eta_c}t}}{ m_{\eta_c}}\Bigg{[} \frac{c_0}{2m_{\eta_c}^2}\Big{(}1+ m_{\eta_c}t -\frac{\eta m_{\eta_c}^2}{Z_0^2\xi}\Big{)} \nonumber \\
&+&\frac{c_1}{m^2_{J/\psi}}\Big{(}\frac{1}{\xi}+\frac{\delta m}{m_{\eta_c}}
+\frac{(\delta m)^2}{2m_{\eta_c}^2}(1+m_{\eta_c}t-\frac{\eta m_{\eta_c}^2}{Z_0^2\xi})\Big{)} \Bigg{]}, \nonumber \\
\eeq
together with
\beq
\mathcal{I}_0(t,0)=\frac{-4e_c Z_0
m_{J/\psi}}{m_{\eta_c}+m_{J/\psi}}\frac{e^{-m_{\eta_c}t}}{ m_{\eta_c}} \left( c_0 +c_1\frac{(\delta m)^2 }{m_{J/\psi}^2}\right), \nonumber \\
\eeq
one can immediately determine $c_0$ and $c_1$ using $H_{\mu\nu}(\vec{x},t)$ as input through
\beq\label{eq:c1}
c_1=\Bigg{[}\tilde{\mathcal{I}}_1(t)-\frac{\xi\tilde{\mathcal{I}}_0(t)}{2m_{\eta_c}^2}\Big{(}1 + m_{\eta_c}t-\frac{\eta m_{\eta_c}^2}{Z_0^2\xi }\Big{)}\Bigg{]}\frac{m_{J/\psi}^2 m_{\eta_c}}{m_{\eta_c}+\xi\delta m} \nonumber \\
\eeq
and
\beq\label{eq:c0}
c_0=\tilde{\mathcal{I}}_0(t)-c_1 \times \frac{(\delta m)^2}{m_{J/\psi}^2},
\eeq
where $\tilde{\mathcal{I}}_n(t)$ are defined as
\beq
\tilde{\mathcal{I}}_n(t)\equiv -\frac{(m_{\eta_c}+m_{J/\psi})}{4e_c Z_{0}m_{J/\psi}}m_{\eta_c}e^{m_{\eta_c}t}\mathcal{I}_n(t,0).
\eeq

\section{Numerical setup}\label{sec:setup}
\begin{table}[!h]
\begin{ruledtabular}
\begin{tabular}{ccccccc}
\textrm{Ens} & $a$ (fm) & $L^3\times T$ & $N_{\textrm{conf}}\times T$
& $m_{\pi} (\textrm{MeV})$ & $t$  & $L$[fm] \\
\hline
a67 & 0.0667(20) & $32^3\times 64$& $197\times 64$ & 300 & 5$-$15 & 2.13 \\
a85 & 0.085(2) & $24^3\times 48$ & $200\times 48$ & 315 & 3$-$12 & 2.04\\
a98 & 0.098(3) & $24^3\times 48$ & $236\times 48$ & 365 & 2$-$11  & 2.35\\
\end{tabular}
\end{ruledtabular}
\caption{
Parameters of gauge ensembles are used in this work. From left to right, we list the ensemble name, the lattice spacing $a$,
the spatial and temporal lattice size $L$ and $T$, the number of the measurements
of the correlation function for each ensemble $N_{\textrm{conf}}\times T$ with $N_{\textrm{conf}}$ the number of the configurations
used, the pion mass $m_{\pi}$, the range of the time separation $t$ between the initial hadron and the electromagnetic current, and the spatial lattice size $L$ in the physical unit.}\label{table:lat_ens}
\end{table}

We use three two-flavor twisted mass gauge ensembles generated by
the Extended Twisted Mass Collaboration (ETMC)~\cite{ETM:2009ptp,Becirevic:2012dc} with lattice spacing
$a \simeq 0.0667,0.085,0.098$ fm. We call these ensembles a67, a85, and a98, respectively. The lattice spacing errors are taken from Ref.~\cite{Blossier:2010cr}, where the lattice spacings are fixed by matching the pion decay constant calculated on the lattice with its physical value, leading to a Sommer parameter $r_0=0.440(12)$ fm. The ensemble parameters are shown in Table.~\ref{table:lat_ens}. The valence charm quark mass is tuned
by setting the lattice result of $J/\psi$ mass to the physical one.
The detailed information on the tuning is referred to in Ref.~\cite{Meng:2021ecs}.

In this work, we calculate the three-point correlation function
$C^{(3)}_{\mu\nu}(\vec{x},t) \equiv\langle \phi_{\eta_c}(\vec{x},t)J_{\nu}(0)\phi_{J/\psi,\mu}^{\dagger}(-t)\rangle$ using $Z_4$-stochastic wall-source $J/\psi$ interpolating operator $\phi_{J/\psi,\mu}=\bar{c}\gamma_{\mu}c$ and point-source $\eta_c$ interpolating operator $\phi_{\eta_c}=\bar{c}\gamma_5c$. In this work, only the connected diagrams are considered. To compute the connected correlation function, we place the wall-source propagator on initial $J/\psi$ and the point source propagator on $\eta_c$ and treat the current as the sink. All the propagators are produced on all time slices by average to increase the statistics based on time translation invariance. The stochastic propagator used here helps to reduce the uncertainty of the mass spectrum by nearly half. We also apply the APE~\cite{APE:1987ehd} and Gaussian smearing~\cite{Gusken:1989qx} to the $J/\psi$ field to efficiently reduce the excited-state effects. The hadronic mass is extracted from the two-point correlation function $C^{(2)}(t)=\langle \mathcal{O}_{h}(t) \mathcal{O}_{h}^{\dagger}(0)\rangle$ using a two-state fit
\be\label{eq:2pt}
C^{(2)}(t)=V \sum_{i=0,1}\frac{(Z_i^{h})^2}{2E_i^{(h)}} \left(\textrm{e}^{-E_i^{(h)}t}+\textrm{e}^{-E_i^{(h)}(T-t)}\right)
\ee
with $V$ the spatial-volume factor, $E_0^{(h)}$ the ground-state energy of the meson $h=m_{\eta_c},m_{J/\psi}$ and $E_1^{(h)}$the energy of the first excited state. $Z_i^{(h)}=\frac{1}{\sqrt{V}}\langle i|\mathcal{O}_{h}^\dagger|0\rangle$ ($i=0,1$) are the overlap amplitudes for the ground and the first excited state. Using $Z_0^{J/\psi}$ and $m_{J/\psi}$ as the inputs, the hadronic function $H_{\mu\nu}(\vec{x},t)$ is then determined through
\be
H_{\mu\nu}(\vec{x},t)=C^{(3)}_{\mu\nu}(\vec{x},t)/[(Z_0^{J/\psi}/2m_{J/\psi})e^{-m_{J/\psi}t}].
\ee

In our calculations, we choose the local vector current $J_\nu(x)=Z_V e_c \bar{c}\gamma_{\nu}c$, where
an additional renormalization factor $Z_V$ is introduced to convert the local vector current to the conserved current, with corrections no larger than $O(a^2)$. The detailed determination of $Z_V$ is provided in our previous work~\cite{Meng:2021ecs}. Here, we use the values $Z_V = 0.6047(19), 0.6257(21)$, and $0.6516(15)$ for $a=0.098,0.085$, and $0.0667$ fm, respectively.

\section{Numerical results}\label{sec:result}

\subsection{Dispersion relation}\label{sec:disper_relation}

\begin{figure}[!h]
\centering
\subfigure{\includegraphics[width=0.48\textwidth]{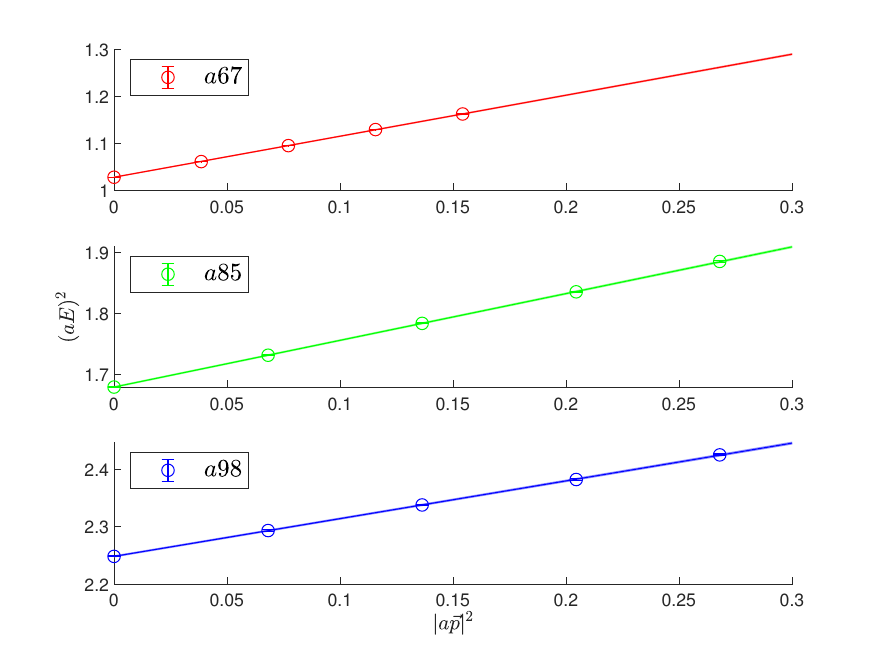}}
\subfigure{\includegraphics[width=0.48\textwidth]{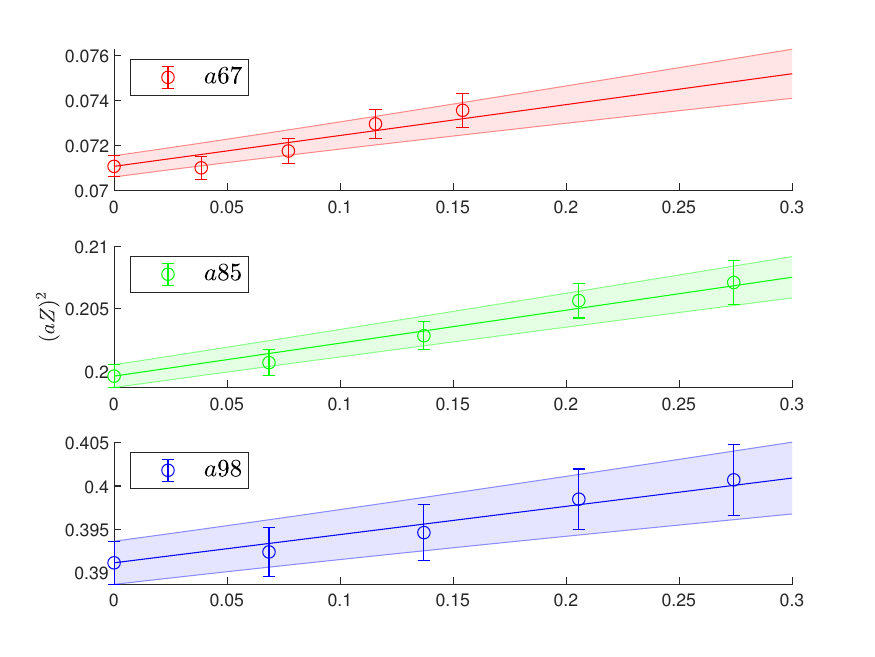}}
\caption{For $\eta_c$ meson, the continuous dispersion relation (top panel) and $Z^2$ as a function of $|a\vec{p}|^2$ (bottom panel), where $\vec{p}=2\pi\vec{n}/L,|\vec{n}|^2=0,1,2,3,4$.}
\label{fig:xi_eta_old}
\end{figure}

In our calculation, the coefficients $\xi$ and $\eta$ introduced in Eq.~(\ref{eq:E_Z}), appear directly in our master formulas.
It is therefore necessary to calculate them first. Both of them can be extracted from the effective levels of the $\eta_c$ particle, which are obtained by fitting the two-point functions in Eq.~(\ref{eq:2pt}).
We calculate $\xi$ and $\eta$ by fitting $E(\vec{p})$ and $Z$ with the formula described in Eq.~(\ref{eq:E_Z}). In Fig.~\ref{fig:xi_eta_old}, lattice results of $E(\vec{p})$ and $Z$ are presented, where five momenta $\vec{p}=2\pi\vec{n}/L$ with $|\vec{n}|^2=0,1,2,3,4$ are considered. It is seen that the simple parametrizations can describe the momentum dependence of $E$ and $Z$ well and there are nice linear behaviors as illustrated. Numerical values of $\xi$ and $\eta$ are summarized in Table~\ref{tab:xi_eta_new}.

\begin{table}[!h]
\begin{ruledtabular}
\begin{tabular}{ccccc}
Ensemble & a67 & a85 & a98 \\
\hline
$\xi$ & 0.8702(32)&0.7658(26) & 0.6558(35)\\
$\eta$ &0.0138(30) & 0.0263(36)&0.0328(88) \\
\end{tabular}
\end{ruledtabular}
\caption{Numerical results of $\xi$ and $\eta$ for all ensembles.}
\label{tab:xi_eta_new}
\end{table}

\subsection{Finite-volume correction (FVC)}\label{sec:FVC}
\begin{figure}[!h]
\centering
\subfigure{\includegraphics[width=0.45\textwidth]{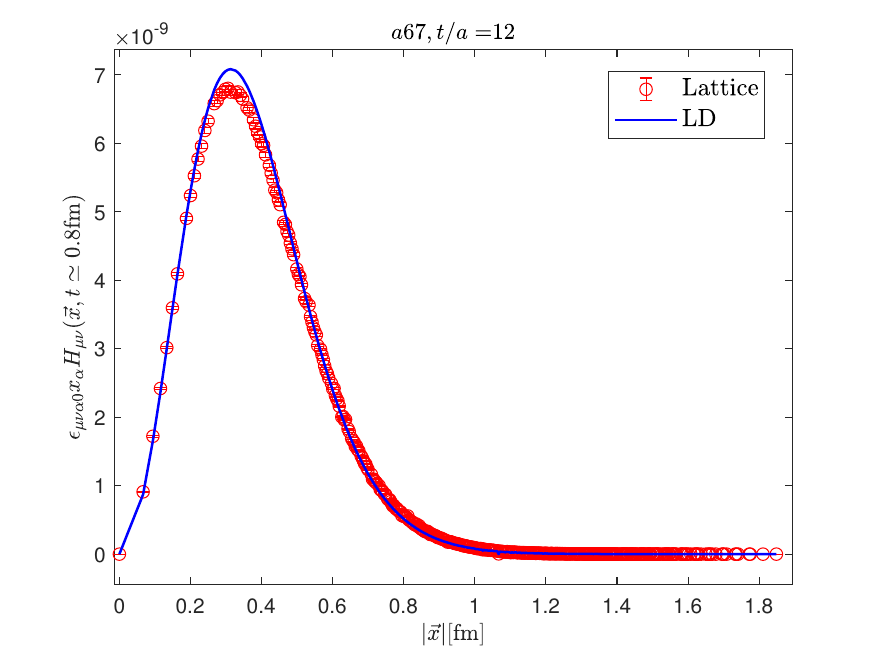}}\hspace{5mm}
\subfigure{\includegraphics[width=0.45\textwidth]{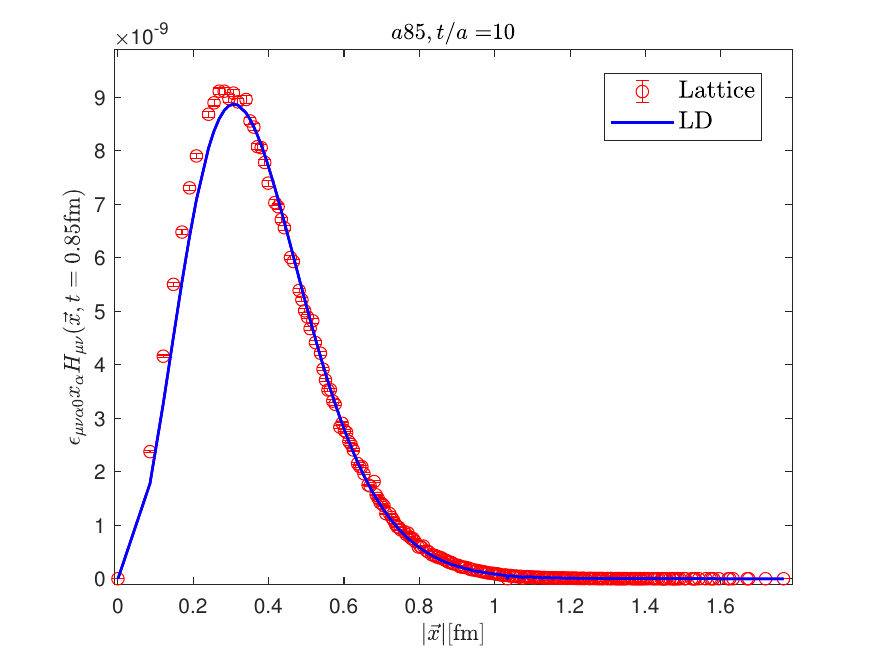}}\hspace{5mm}
\subfigure{\includegraphics[width=0.45\textwidth]{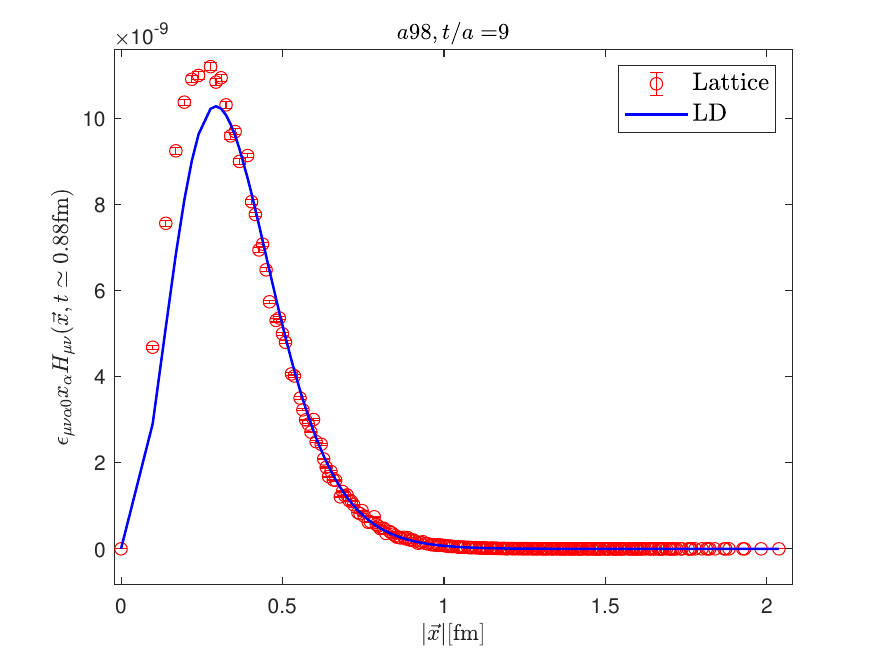}}\hspace{5mm}
\caption{For all ensembles, $\epsilon_{\mu\nu\alpha 0}x_{\alpha}H_{\mu\nu}(\vec{x},t)$ as a function of $|\vec{x}|$ are given. The results of the direct lattice calculation (lattice) and constructed long-distance version (LD) are presented, respectively.}
\label{fig:LD}
\end{figure}

The finite-volume effects should only have a tiny contribution 
because $H_{\mu\nu}(\vec{x},t)$ is dominated by $\eta_c$ state, it is exponentially suppressed when $|\vec{x}|$ becomes large. However, for the quantity $|\vec{x}|^2\epsilon_{\mu\nu\alpha 0}x_{\alpha}H_{\mu\nu}(\vec{x},t)$ that contains a factor $|\vec{x}|^2$ that increases rapidly with distance, these finite-volume effects may have a non-negligible contribution. To estimate these effects, we construct a long-distance hadronic function $H_{\mu\nu}^{LD}(\vec{x},t)$
\beq\label{eq:H_model}
H_{\mu\nu}^{LD}(\vec{x},t)&\doteq&\frac{-2e_cm_{J/\psi}}{m_{\eta_c}+m_{J/\psi}}\frac{1}{L^3}\sum\limits_{\vec{p}}Z(\vec{p})V(q^2)  \nonumber \\
&\times&\epsilon_{\mu\nu\alpha 0}p_{\alpha}\sin(\vec{p}\cdot \vec{x})\frac{e^{-Et}}{E}
\eeq
with the transition factor $V(q^2)=d_0/(1-d_1q^2)$ taken into account. $d_0$ and $d_1$ are input parameters that can be determined by matching the long-distance hadronic function $H_{\mu\nu}(\vec{x},t)$ and $H_{\mu\nu}^{LD}(\vec{x},t)$ for a sufficiently large $t$. In Fig.~\ref{fig:LD}, such matchings are presented and the lattice data are well described by the model with the coefficient $d_0=1.66,1.55,1.40$ and $d_1=-1.65,-1.18,-0.9$ for
a67, a85, and a98, respectively. For the necessity and the detailed study of the FVC, see related discussion in Appendix~\ref{sec:appendixA} and \ref{sec:appendixB}. Then, the space integral in Eq.~(\ref{eq:I_0}) and Eq.~(\ref{eq:I_1}) can be divided into two parts, i.e. inside the box $\int_V$ and outside the box $\int_{>V}$. The scalar functions $\mathcal{I}_0$ and $\mathcal{I}_1$ are obtained by
\beq\label{eq:I0_FVC}
\mathcal{I}_0(t,|\vec{p}|)&=&\int_{V} d^3\vec{x}\frac{j_1(|\vec{p}||\vec{x}|)}{|\vec{p}||\vec{x}|}\epsilon_{\mu\nu \alpha 0}x_{\alpha}H_{\mu\nu}(\vec{x},t) \nonumber \\
&+&\int_{>V} d^3\vec{x}\frac{j_1(|\vec{p}||\vec{x}|)}{|\vec{p}||\vec{x}|}\epsilon_{\mu\nu \alpha 0}x_{\alpha}H_{\mu\nu}^{LD}(\vec{x},t),
\eeq
and
\beq\label{eq:I1_FVC}
\mathcal{I}_1(t,0)&=&\frac{1}{30}\int_V d^3\vec{x}|\vec{x}|^2\epsilon_{\mu\nu\alpha 0}x_{\alpha}H_{\mu\nu}(\vec{x},t) \nonumber \\
&+&\frac{1}{30}\int_{>V} d^3\vec{x}|\vec{x}|^2\epsilon_{\mu\nu\alpha 0}x_{\alpha}H_{\mu\nu}^{LD}(\vec{x},t),
\eeq
where $H_{\mu\nu}(\vec{x},t)$ denotes the hadronic function calculated on the lattice directly. Both the second terms give the estimation of the 
finite-volume effects.

\subsection{Transition factor}\label{sec:form_factor}
\begin{figure}[!h]
\centering
\subfigure{\includegraphics[width=0.48\textwidth]{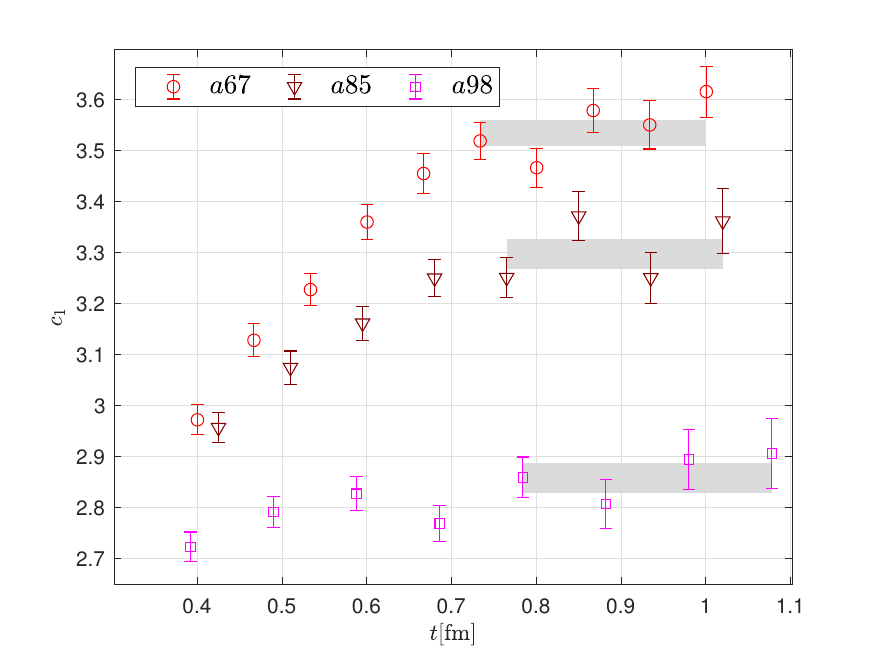}}\hspace{5mm}
\subfigure{\includegraphics[width=0.48\textwidth]{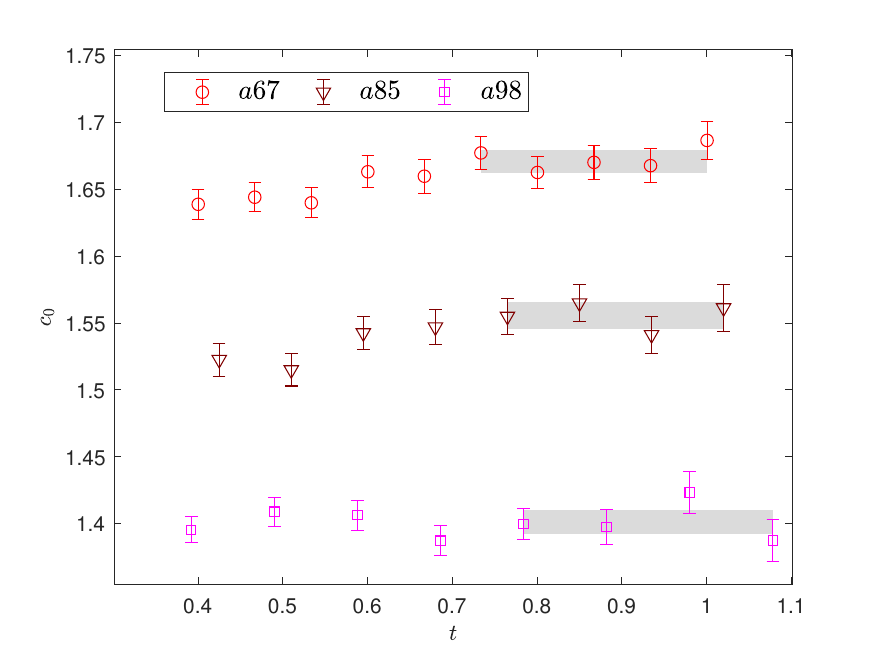}}\hspace{5mm}
\caption{Lattice results of $c_1$ and $c_0$ as a function of $t$. The $c_1$ and $c_0$
are calculated using Eq.~(\ref{eq:c1}) and Eq.~(\ref{eq:c0}), respectively.}
\label{fig:c01_FVC}
\end{figure}

The lattice results of $c_0$ and $c_1$ as a function of the time separation $t$ are shown
in Fig.~\ref{fig:c01_FVC}.
We perform the correlated fit of them to a constant at large $t$, and
determine them. The corresponding values are listed in Table.~\ref{tab:c01_FVC}.

\begin{figure}[!h]
\centering
\subfigure{\includegraphics[width=0.48\textwidth]{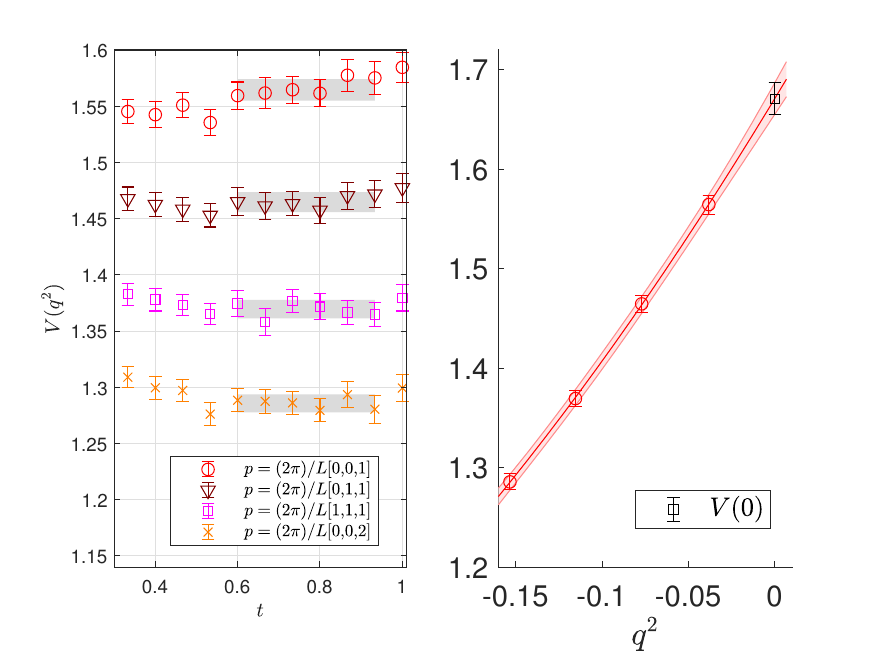}}\hspace{5mm}
\caption{Using a67 as an example, $V(q^2)$ as a function of $t$ with different momenta is shown
in the left panel and the continuous extrapolation in $q^2$ is shown in the right panel.}
\label{fig:Vq2_old}
\end{figure}

The traditional approach is also adopted for further comparison with the new method, where the $V(q^2)$ is given in the following way
\beq\label{eq:Vq2_old}
V(q^2)&=&-\frac{(m_{\eta_c}+m_{J/\psi})}{4e_cZm_{J/\psi}}Ee^{Et}
\nonumber \\
&\times&\frac{i}{|\vec{p}|^2}\epsilon_{\mu\nu\alpha0}p_{\alpha}\int d^3\vec{x}e^{-i\vec{p}\cdot \vec{x}}H_{\mu\nu}(\vec{x},t)
\eeq
with $\vec{p}=\frac{2\pi}{L}\vec{n}$ for $\vec{n}=(0,0,1),(0,1,1),(1,1,1)$, and $(0,0,2)$.

The results of $V(q^2)$ as a function of $t$ are shown in the left panel of Fig.~\ref{fig:Vq2_old}, together with
the $q^2$ dependence in the right panel. Here we take the finest ensemble a67 as an
example. The continuous extrapolation in $q^2$ is performed by a polynomial function
$V(q^2)=c_0+c_1q^2/m_{J/\psi}^2+c_2q^4/m_{J/\psi}^4$.
The $c_i$ is then obtained by a correlated fit of the lattice data
and the results are summarized in Table.~\ref{tab:c01_FVC}. These results are consistent with the ones from
the new method, but the errors of $c_0$ are 1.7$-$1.8 times larger and $c_1$ even an order of magnitude larger.

\begin{table}[!h]
\begin{ruledtabular}
\begin{tabular}{ccc|ccc}
& \multicolumn{2}{c}{New}& \multicolumn{3}{c}{Traditional} \\
\cmidrule(r){2-6}
 Ensemble & $c_0$ & $c_1$ & $c_0$ & $c_1$ & $c_2$ \\

\hline
$a67$ & 1.670(09) &3.53(3) & 1.671(16) & 3.16(33) & 2.8(1.8) \\
$a85$ & 1.556(10) &3.30(3) & 1.547(17) & 2.78(37) & 2.1(1.9) \\
$a98$ & 1.401(09) &2.86(3) & 1.366(16) & 1.44(45) & $-$5.5(3.1) \\
\end{tabular}
\end{ruledtabular}
\caption{Numerical results of $c_0$ and $c_1$ for all ensembles. The $c_2$ from the traditional method is also presented. A polynomial function of $V(q^2)$ is adopted in the traditional approach. A change in the sign of $c_2$ for a98 implies that it may have a certain model dependence for this ensemble.}
\label{tab:c01_FVC}
\end{table}

\begin{figure}[!h]
\centering
\subfigure{\includegraphics[width=0.48\textwidth]{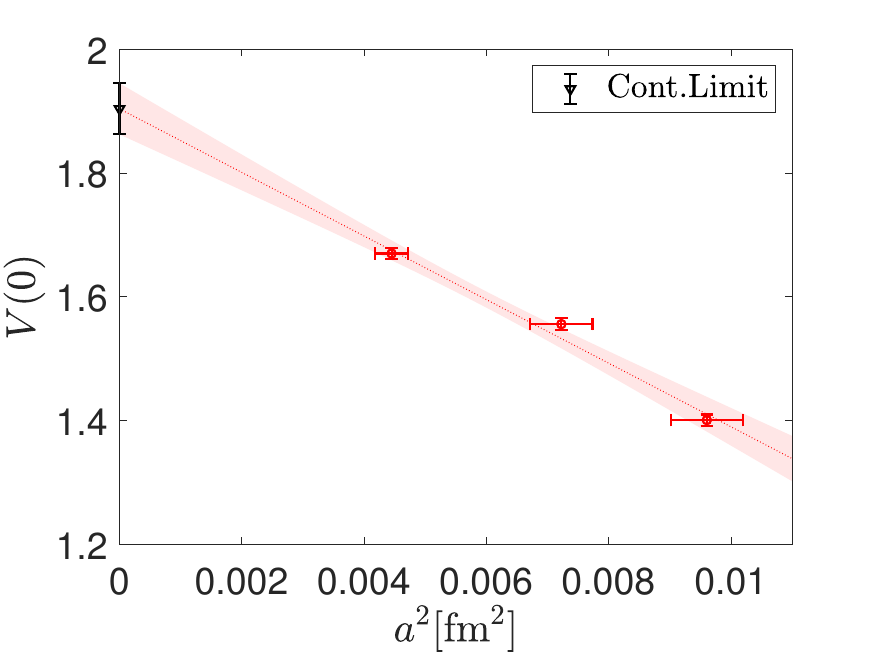}}\hspace{5mm}
\caption{Lattice results of $V(0)$ as a function of lattice spacing. The errors of lattice spacing are included
in the fitting and presented by the horizontal error bars. The symbol of the red circle denotes the lattice results
from ensemble a67, a85, and a98 from left to right. The black triangle is the result in the continuous limit $a^2\rw 0$.}
\label{fig:cont_limit}
\end{figure}

In Fig.~\ref{fig:cont_limit}, an extrapolation of $V(0)$ in linear $a^2$ is performed under three
different lattice spacings. Such behavior is expected for the twisted mass configuration which is
automatically $O(a)$ improved. The fit describes the lattice data well, indicating that no ensemble we utilized has a residual $\mathcal{O}(a)$ effect. This conclusion has also been demonstrated in previous study~\cite{Becirevic:2012dc} and in other physical quantities~\cite{ETM:2009ptp,Alexandrou:2009qu,ETM:2009ztk,Meng:2021ecs}. Finally, we report the determination of the on shell form factor as
\be
V(0)=1.90(4),
\ee
where the error is the statistical error obtained with the lattice spacing errors taken into account.

Using the on shell transition factor $V(0)$ as input, one can determine the partial decay width of $J/\psi\rightarrow \gamma\eta_c$ by
\beq
\Gamma(J/\psi\rightarrow \gamma\eta_c)&=&\alpha e_c^2\frac{16}{3}\frac{|\vec{k}|^3} {(m_{J/\psi}+m_{\eta_c})^2}|V(0)|^2 \nonumber \\
&=& 2.30(10)~\textrm{keV},
\eeq
where $\alpha=1/137.036$~\cite{PDG24} and $|\vec{k}|=(m_{J/\psi}^2-m_{\eta_c}^2)/(2m_{J/\psi})$. $m_{\eta_c}$ and $m_{J/\psi}$ are the experimental values quoted by PDG~\cite{PDG24}. Then, the branching fraction is obtained as
\be
\operatorname{Br}(J/\psi\rightarrow \gamma\eta_c)=2.49(11)_{\textrm{lat}}(5)_{\textrm{exp}}\%
\ee 
where the first error is the statistical error from the lattice simulation, and the second comes from the uncertainty of $J/\psi$ total decay width $92.6(1.7)$ keV.

In our previous work, the two-photon decay width of $\eta_c$ particle is calculated by $\Gamma(\eta_c\rightarrow 2\gamma)=6.67(16)(6)$ keV~\cite{Meng:2021ecs}. Combining with the decay width $\Gamma(J/\psi \rightarrow \gamma \eta_c)=2.30(10)$ keV and taking into account $\eta_c$ total decay width $\Gamma_{\eta_c}^{\textrm{total}}=30.5(0.5)$ MeV~\cite{PDG24}, we finally obtain the product branching fraction $\operatorname{Br}(J/\psi\rightarrow \gamma \eta_c)\times \operatorname{Br}(\eta_c\rightarrow 2\gamma)=5.43(42)\times 10^{-6}$. While the paper is under review, the most recent BESIII collaboration reports a determination of the product branching fraction $\operatorname{Br}(J/\psi\rightarrow \gamma \eta_c)\times \operatorname{Br}(\eta_c\rightarrow 2\gamma)=5.23(26)(30)\times 10^{-6}$~\cite{BESIII:2024rex}, which is well consistent with our lattice calculation.
 
\section{Discussion}\label{sec:discussion}
 
Our lattice result is consistent with previous lattice studies~\cite{Gui:2019dtm,Becirevic:2012dc,Donald:2012ga,Delaney:2023fsc,Colquhoun:2023zbc} as shown in Fig.~\ref{fig:Jpsi_etac_others}. We only collect the lattice results that use at least three different lattice spacings for 
a continuum limit. The latest PDG have updated the branching ratio of $J/\psi\rightarrow \gamma\eta_c$ as $1.41(14)\%$, and the corresponding on shell transition factor is $V_{\textrm{PDG}}(0)=1.43(7)$. The PDG value is also plotted in Fig.~\ref{fig:Jpsi_etac_others} for a better comparison with the lattice calculations. We currently ignore the effects of the disconnected diagrams for convenience to test the validity of the method. Besides, the quenching of strange and charm quarks, the unphysical pion masses and finite-volume effect may also account for the difference for our calculation. However, a systematic study by HPQCD has found these effects only have small contributions~\cite{Colquhoun:2023zbc}. At present, it is still not clear what accounts for the difference between the lattice and the experiment. To clarify this discrepancy, it is crucial to estimate the contribution of disconnected diagrams in the future, which has never been studied to date. Compared with previous approaches, our method has the following peculiar advantages:

\begin{figure}[!h]
\centering
\subfigure{\includegraphics[width=0.48\textwidth]{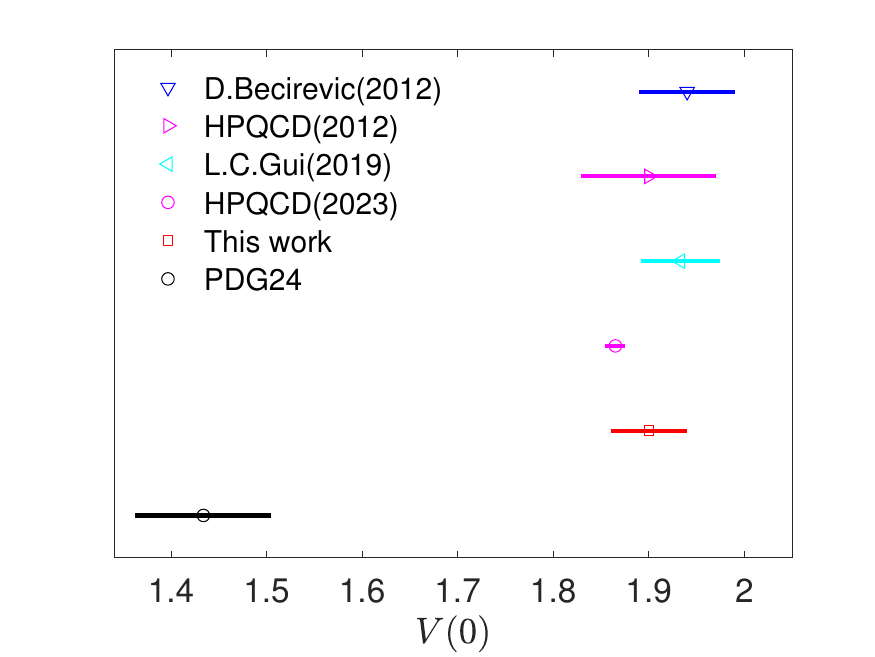}}\hspace{5mm}
\caption{Results of the on shell transition factor $V(0)$ by lattice QCD using at least three different lattice spacings. The value of PDG24 is also listed for a comparison which is denoted by a black circle.}
\label{fig:Jpsi_etac_others}
\end{figure}

\bit

\item[i)] The on shell transition factor is simply extracted from the hadronic function $H_{\mu\nu}(\vec{x},t)$,
which can be calculated on the lattice directly. The method has no dependence on the modeling of the
momentum extrapolation for the off shell transition factors, thus avoiding an additional model-dependent uncertainty. The statistical errors of the on-shell transition factor from the new method are reduced by 40\%, while the errors of their slopes are even an order of
magnitude smaller.

\item[ii)] Our calculations are completed only with the generation of point-source and wall-source propagators.
The propagators generated in this project can be generally applied to other charmonium decays,
like $\eta_c/\chi_{c_0}\rw 2\gamma$~\cite{CLQCD:2020njc,Meng:2021ecs,Zou:2021mgf}, $J/\psi \rw 3\gamma$~\cite{Meng:2019lkt}, and $J/\psi \rw \gamma \nu\bar{\nu}$~\cite{Meng:2023bjc}. Hence, the method developed here sheds light on the precision determination of the disconnected diagrams in these processes.

\eit

\section{Conclusion}\label{sec:conclude}
In this study, we propose a model-independent method to study the radiative transition process and apply it to the charmonium transition decay $J/\psi \rw \gamma\eta_c$.
This method enables direct determination of the on shell transition factor from the lattice hadronic function.
As a result, we obtain the on shell transition factor for $J/\psi\rw \gamma\eta_c$ as
$V(0)=1.90(4)$ after a controlled continuous limit, which is well consistent with previous lattice studies.

The method eliminates the need for momentum extrapolation, thereby avoiding possible model dependencies of the
transition factors. As it does not require additional quark propagators
on twisted boundary conditions, the propagators we generated in this study can also be utilized for other charmonium studies. This is particularly valuable for lattice calculations involving the disconnected diagrams,
which are much more computationally intensive, e.g., the charmonium mass determinations~\cite{McNeile:2004wu,deForcrand:2004ia,Levkova:2010ft,Zhang:2021xrs} and radiative decays~\cite{Jiang:2022gnd}.

\begin{acknowledgments}
We thank ETM Collaboration for sharing the gauge configurations with us. Y.M. is grateful to Xu Feng for very valuable discussions and comments on reading through the manuscripts.
The authors acknowledge the support from the NSFC of China under the Grant 
No.12293060, 12293063, 12305094, 11935017.
The calculation was carried out on the Tianhe-1A supercomputer at Tianjin National Supercomputing Center and the SongShan supercomputer at the National Supercomputing Center in Zhengzhou.
\end{acknowledgments}.

\bibliographystyle{apsrev4-2}

\bibliography{ref}
\clearpage

\begin{widetext}
\appendix

\section{Lattice results of $c_0,c_1$ without FVC}
\label{sec:appendixA}
The results of $c_0$ and $c_1$ with no FVC included are plotted in Fig.~\ref{fig:c01}. It is seen that $c_1$ has a relatively unstable plateau, especially for the ensembles with small volumes, \textit{i.e.} a67 and a85. Numerical values of $c_0$ and $c_1$ are summarized in Tab.~\ref{tab:c01}. The FVC causes
$c_1$ to deviate by 7\%, 9\% and 4\% for a67, a85, and a98, respectively. However, the influence of $c_1$ on $c_0$ is highly suppressed by the factor
$(\delta m)^2/m_{J/\psi}^2$, resulting in a change in $c_0$ of no more than 1\%. In a continuum limit, we obtain the $c_0$ as $1.89(4)$. The value is consistent with $1.90(4)$, where the FVC is included. 

\begin{figure}[!h]
\centering
\subfigure{\includegraphics[width=0.48\textwidth]{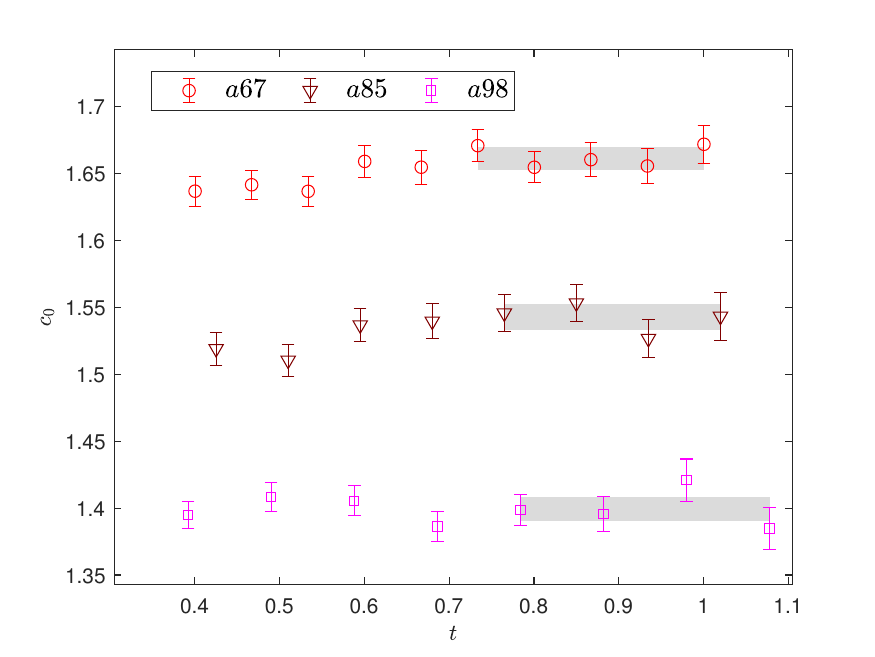}}\hspace{5mm}
\subfigure{\includegraphics[width=0.48\textwidth]{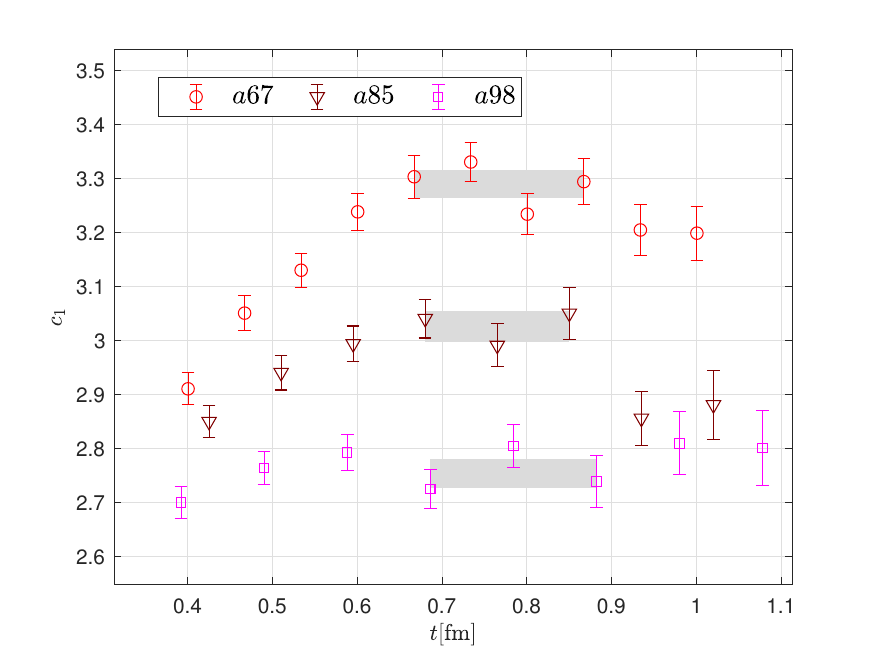}}\hspace{5mm}
\caption{Lattice results of $c_1$ and $c_0$ without including the FVC.}
\label{fig:c01}
\end{figure}

\begin{table}[!h]
\begin{ruledtabular}
\begin{tabular}{ccccc}
Ensemble & a67 & a85 & a98  & Cont.Limit \\
\hline
$c_0$ & 1.660(09)& 1.546(10) & 1.393(08)& 1.89(4) \\
$c_1$ & 3.29(3)  & 3.03(3)& 2.75(3) & ---\\
\end{tabular}
\end{ruledtabular}
\caption{Numerical results of $c_0$ and $c_1$ without the finite-volume corrections included, together with the continuum limit of $c_0$.}
\label{tab:c01}
\end{table}

\section{$L-$dependence of FVC}
\label{sec:appendixB}
In this section, we give the numerical verification that the finite-volume corrections in Eq.~(\ref{eq:I0_FVC}) and Eq.~(\ref{eq:I1_FVC}) are exponentially suppressed as the volume size increases. To that end, we introduce the following ratios
\beq
R(L)&\equiv& \frac{\int_{L^3}^{\infty}\epsilon_{\mu\nu\alpha 0}x_{\alpha}H_{\mu\nu}^{LD}(\vec{x},t)}{\int_0^{\infty}\epsilon_{\mu\nu\alpha 0}x_{\alpha}H_{\mu\nu}^{LD}(\vec{x},t)} \\
R_2(L)&\equiv& \frac{\int_{L^3}^{\infty}|\vec{x}|^2\epsilon_{\mu\nu\alpha 0}x_{\alpha}H_{\mu\nu}^{LD}(\vec{x},t)}{\int_0^{\infty}|\vec{x}|^2\epsilon_{\mu\nu\alpha 0}x_{\alpha}H_{\mu\nu}^{LD}(\vec{x},t)}
\eeq
which characterize the $L-$dependence of the FVC relative to the total contribution for the scalar quantity $\epsilon_{\mu\nu\alpha 0}x_{\alpha}H_{\mu\nu}^{LD}(\vec{x},t)$ and $|\vec{x}|^2\epsilon_{\mu\nu\alpha 0}x_{\alpha}H_{\mu\nu}^{LD}(\vec{x},t)$.
In Fig.~\ref{fig:R_L}, $R(L)$ and $R_2(L)$ as a function of volume size $L$ are plotted. It shows that $R$ and $R_2$ decrease rapidly as $L$ increases. Besides, $|\vec{x}|^2\epsilon_{\mu\nu\alpha 0}x_{\alpha}H_{\mu\nu}^{LD}(\vec{x},t)$ is more significantly affected by FVC because of the factor $|\vec{x}|^2$ compared to $\epsilon_{\mu\nu\alpha 0}x_{\alpha}H_{\mu\nu}^{LD}(\vec{x},t)$. It is also concluded the FVC of $c_1$ can be completely ignored when the lattice size is larger than 3 fm, and then one can calculate the $c_1$ straightforwardly using the Eq.~(\ref{eq:I_0}) and Eq.~(\ref{eq:I_1}).

\begin{figure}[!h]
\centering
\subfigure{\includegraphics[width=0.48\textwidth]{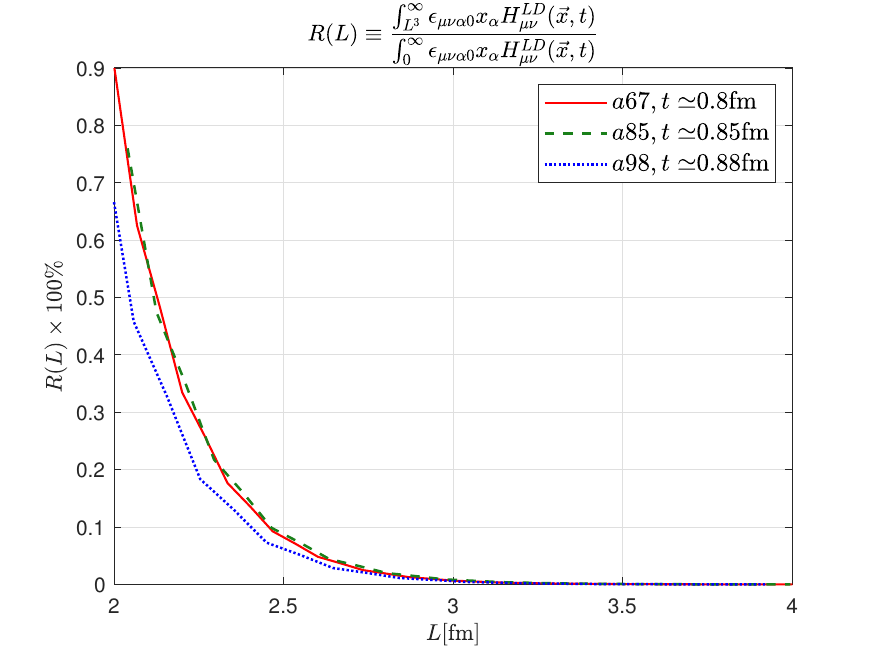}}\hspace{5mm}
\subfigure{\includegraphics[width=0.48\textwidth]{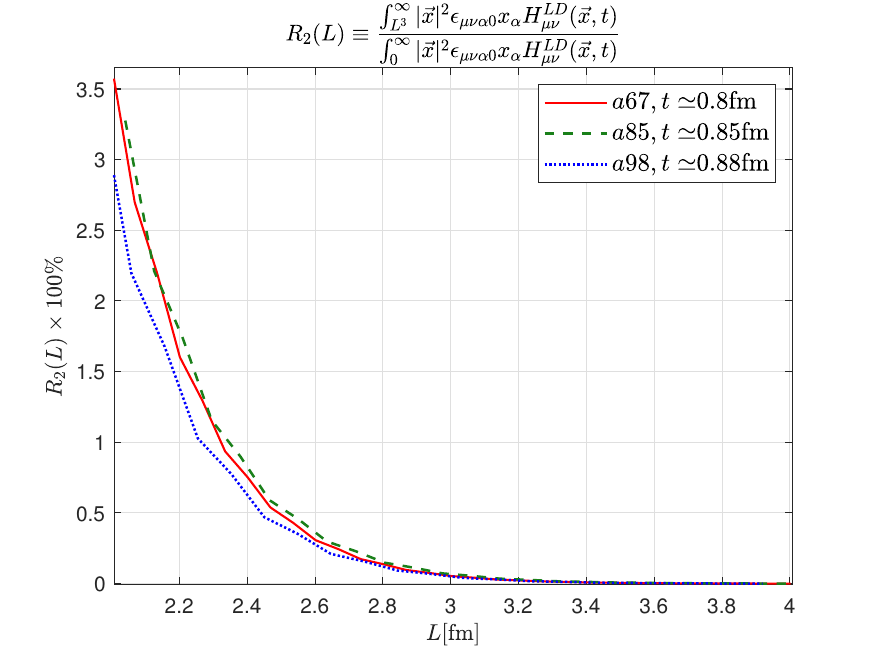}}\hspace{5mm}
\caption{$L-$dependence of the ratio $R(L)$ and $R_2(L)$ for all ensembles.}
\label{fig:R_L}
\end{figure}

\end{widetext}
\end{document}